\documentclass[twocolumn]{aastex62}

\usepackage{color}

\shorttitle{Old star cluster dynamical clock}
\shortauthors{Andr\'es E. Piatti}

\begin{document}

\title{Dependence of the old star clusters' dynamical clock on the host galaxy gravitational field}

\author[0000-0002-8679-0589]{Andr\'es E. Piatti}
\affiliation{Instituto Interdisciplinario de Ciencias B\'asicas (ICB), CONICET-UNCUYO, Padre J. Contreras 1300, M5502JMA, Mendoza, Argentina}
\affiliation{Consejo Nacional de Investigaciones Cient\'{\i}ficas y T\'ecnicas (CONICET), Godoy Cruz 2290, C1425FQB,  Buenos Aires, Argentina}
\correspondingauthor{Andr\'es E. Piatti}
\email{e-mail: andres.piatti@unc.edu.ar}

\begin{abstract}
I report outcomes of the analysis of the $A^+$ parameter, which measures 
the level of radial segregation of blue straggler stars in old star clusters, 
commonly known as the dynamical clock for the long-term internal dynamical evolution.
I used $A^+$ values available in the literature for 48 Milky Way globular clusters. 
I found that the relationship of 
$A^+$ and  the number of central relaxation times which have elapsed
($N_{relax}$) shows a non negligible dependence 
on the strength of the host galaxy
gravitational potential, in addition to depending on the two-body relaxation mechanism. 
Indeed, a measured $A^+$ value corresponds to relatively smaller or larger $N_{relax}$
values for star clusters located farther or closer to the galaxy center.
From an observational point of view, this finding reveals the possibility
of disentangling for the first time 
the dynamical evolutionary stage due to two-body relaxation and tidal effect, 
that affect the whole star 
clusters' body concurrently.
\end{abstract}


\section{Introduction}

The internal dynamics of star clusters has long been addressed in the literature
from numerical and observational studies  \citep{mh1997,hh03,krauseetal2020}. Recently, 
\citet[][see also references therein]{ferraroetal2018}
found a strong correlation between $A^+$, defined as the area enclosed between 
the cumulative radial distribution of blue straggler stars and that of a reference population,
and the number of central relaxation times ($N_{relax}$=age/$t_{rc}$) of old star
clusters.  
Because of the observed correlation between $A^+$ and 
the central relaxation time of old star clusters,  $A^+$ has been used
 as a powerful dynamical clock. 
For the sake of the reader, I refer to a review by \citet{ferraroetal2020}. 

Because $A^+$ measures the overall internal dynamical stage of an old star cluster, 
the evolutionary stage
due to two-body relaxation and tidal effects are both included in the $A^+$ values.  
Particularly, tidal
effects accelerate mass segregation and two-body relaxation by increasing the mass loss rate. 
As far as I am aware, the impact of tidal effects on $A^+$
has not been explicitly mentioned.
The novelty of this
work consists in disentangling observationally, for the first time, the dependence of
$A^+$ on two-body relaxation and tidal effects, so that the dynamical evolutionary stage of a star cluster
due to two-body relaxation can be estimated for star clusters in different orbits. 
In doing this, I  evaluate the contribution of tidal effects to the measured values 
of $A^+$.

\begin{figure*}
\includegraphics[width=0.35\textwidth]{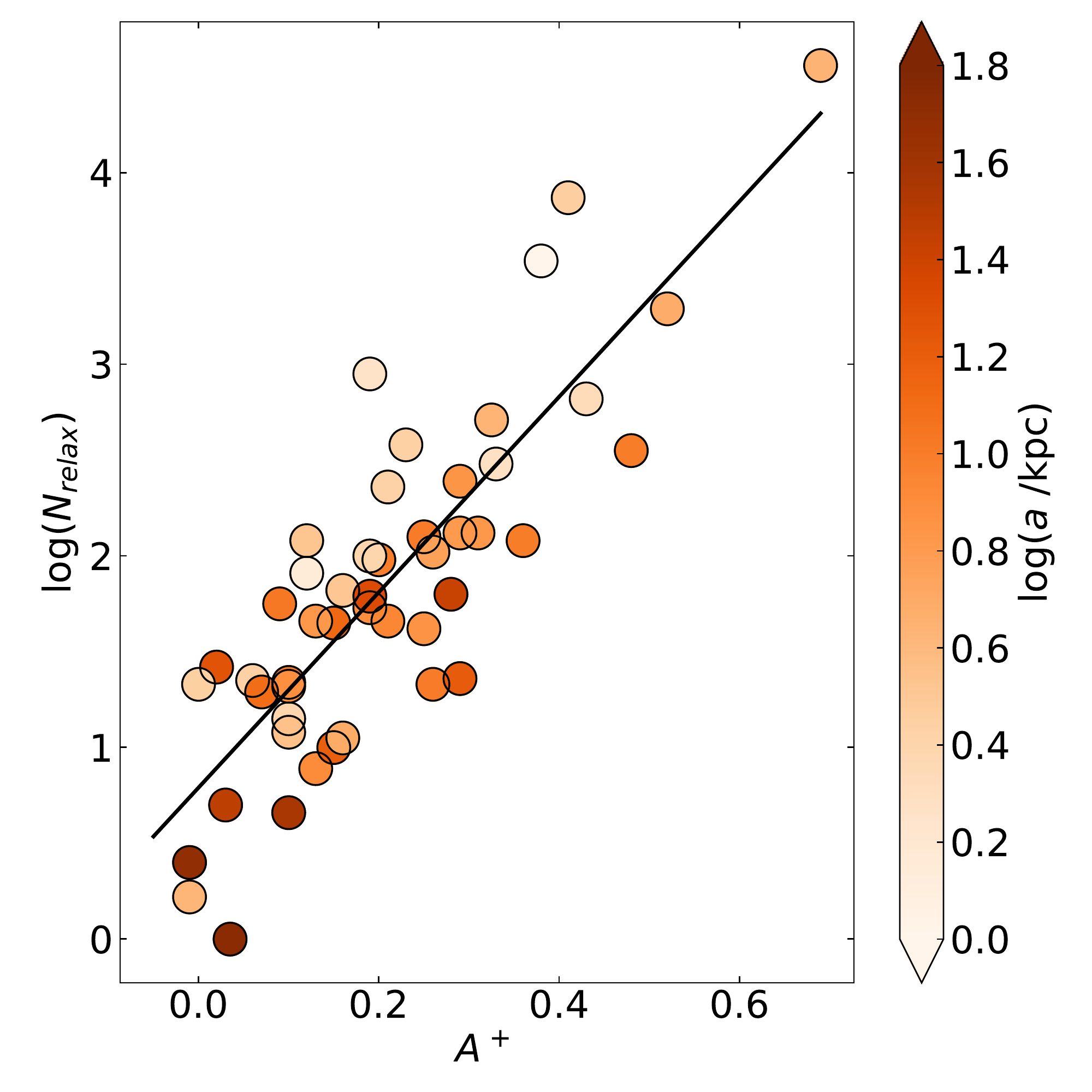}
\includegraphics[width=0.35\textwidth]{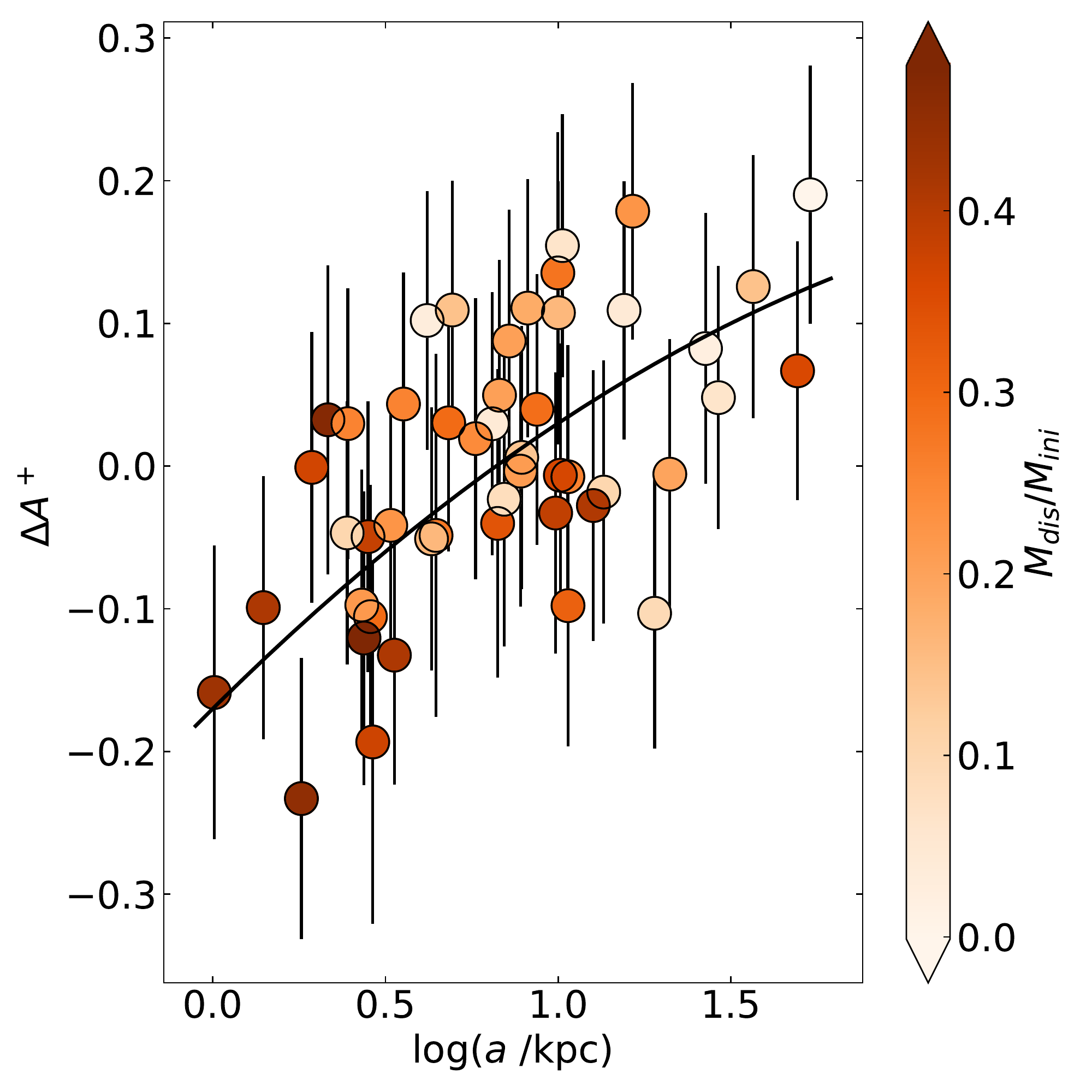}
\includegraphics[width=0.35\textwidth]{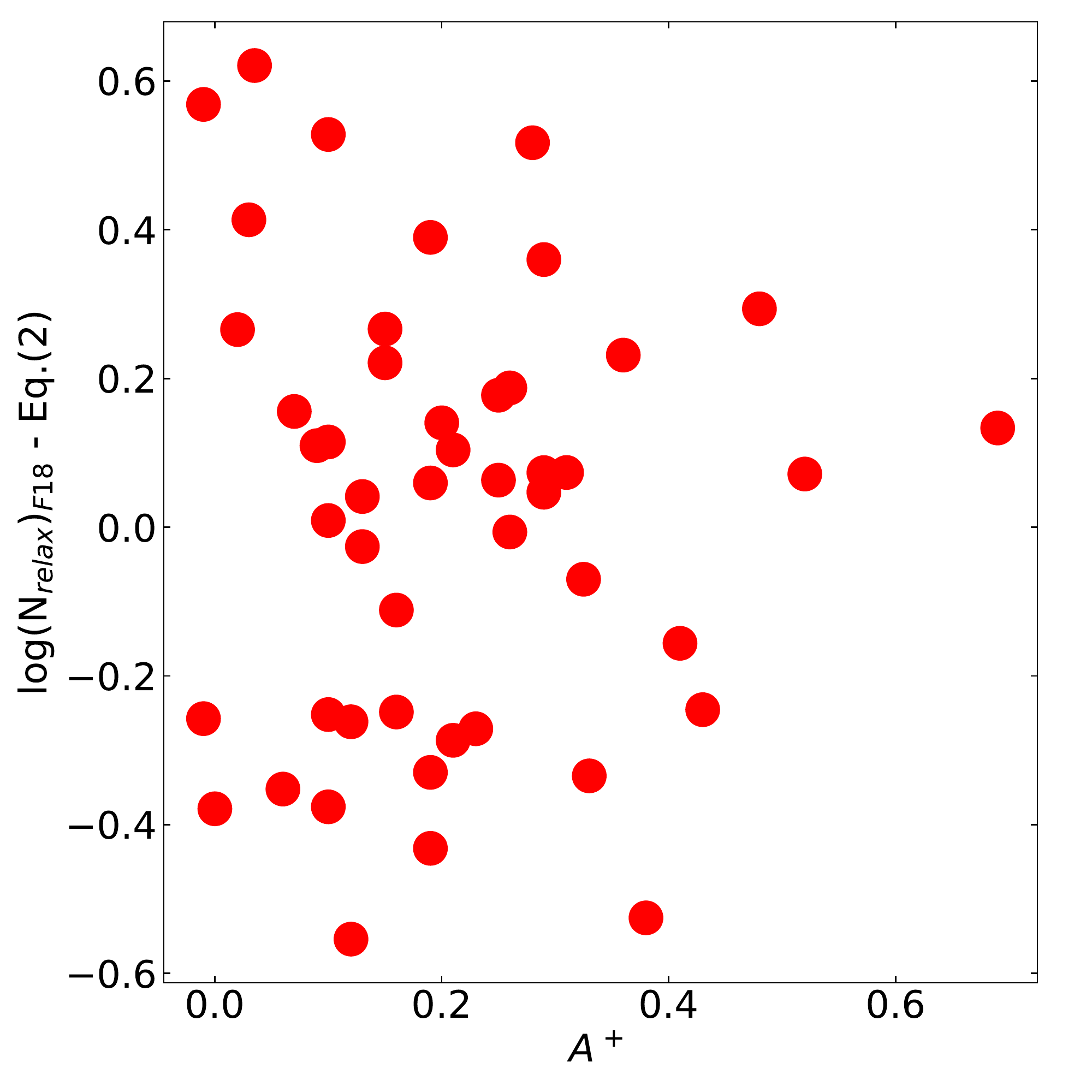}
\caption{{\it Left: }Relation between $A^+$ and log($N_{relax}$) for the 48 Galactic globular clusters
analyzed by \citet{ferraroetal2018}, and their derived least square fit drawn with a solid black line.
 {\it Middle:} $\Delta$$A^+$ as a function of the semi-major axis. Error bars are included. The solid line represents eq.(1).
{\it Right:} Difference between $N_{relax}$ values calculated from \citet{ferraroetal2018}'s
relation and from eq. (2).}
\label{fig1}
\end{figure*}

\section{Analysis and discussion}

Figure~\ref{fig1} (left panel) reproduces the relation obtained by \citet{ferraroetal2018}. They have been colored according to the semi-major axis ($a$) of the cluster's orbits around the Galactic center. I use $a$  because is more representative of  the mean orbital distance of the globular 
clusters than the perigalactic and apogalactic distances  \citep{piatti2019}. 
  A not subtle dispersion in log($N_{relax}$) at a constant $A^+$ 
value is observed, 
which would seem to change with $a$, in the sense that the larger the log($N_{relax}$) value, the 
smaller the mean semi-major  axis. Such a trend, observed in Figure~\ref{fig1}, 
reveals  the correlation with the semi-major
axis or, in other words, that depending on the position in the Milky Way, $A^+$
corresponds to slightly different internal dynamic evolutionary stages.

In order to show the effects of tides in the \citet{ferraroetal2018}'s relation we
evaluate the range of  $A^+$  by computing the 
difference between   the $A^+$ values and those located on the solid line 
of Figure~\ref{fig1} for 
the same $N_{relax}$ values, called  $\Delta$$A^+$.  
The result, depicted in  Figure~\ref{fig1} (middle panel), shows the correlation 
of $\Delta$$A^+$ with the positions of the globular clusters in the Milky Way. 
Points have been colored according to the ratio between the cluster mass lost
by tidal disruption and the initial cluster mass ($M_{dis}/M_{ini}$ ), which we
use as the indicator for tidal field strength, following \citet{piattietal2019b}'s
results. $M_{dis}/M_{ini}$ and $a$ are interchangeable. Uncertainties
in $\Delta$$A^+$ were computing by adding in quadrature those 
of $A^+$ values and the rms  error  of the solid line derived by  \citet{ferraroetal2018}. 
 Figure~\ref{fig1} (middle panel) also shows that $\Delta A^+$ $\approx$
 0.0 dex at the Sun's galactocentric distance. 

Figure~\ref{fig1} (middle panel) reveals that the scatter of points seen in Fig.~\ref{fig1} (left panel)  on both 
sides of the straight line - i.e., the dispersion of $A^+$ values at a fixed 
$N_{relax}$ - is
not random, but caused by the dependence of $A^+$ on $a$. $A^+$ is
larger than the mean $A^+$ value for the corresponding $N_{relax}$ for globular clusters
located toward the outer Milky Way regions, and is smaller than that mean $A^+$ value 
for globular clusters located toward the inner Milky Way regions. This behavior of $A^+$
is expected, because $A^+$ is a measure of the level of spatial segregation of blue
straggler stars. Similarly to other globular cluster stellar populations, they differentially 
experience the effects of the Milky Way gravitational field, so that they can more easily
reach larger distance from the globular cluster center when the Galactic potential well is
weaker, and role reversal. Therefore, parameters that measure the spatial distribution
of stellar populations of globular clusters (e.g., core, tidal radii, relaxation time, etc)
reflect both the internal dynamical evolutionary stage of a globular cluster and the tidal
effects, concurrently. From a purely stellar dynamical point of view, a globular
cluster located in the inner Milky Way region would appear dynamically accelerated,
or in a more advanced internal dynamical evolutionary stage compared to a globular
cluster located in the outer Milky Way regions \citep[see][]{pm2018,piattietal2019b}. 
For completeness purposes I  performed 
a quadratic least square fit between $\Delta$$A^+$ and log($a$ /kpc), as
follows:

\begin{equation}
\Delta A^+ = C_0 + C_1 \times log(a /kpc) + C_2 \times (log(a /kpc))^2.
\end{equation}

\noindent I obtained $C_0$ = -0.17$\pm$0.05, $C_1$ = 0.24$\pm$0.12, and $C_2$ = 
-0.04$\pm$0.07, rms = 0.07, $\chi^2$ = 0.65, and a correlation coefficient = 0.60 (see
solid line in Fig.~\ref{fig1}, middle panel). 
I note that a linear least square fit would
produce similar results with a slightly smaller correlation coefficient.

Eq. (1) can be useful to compute dynamical ages of a star clusters
due to two-body relaxation, as it were not affected by tidal effects. In doing
this, the calculated $\Delta$$A^+$ value must be subtracted from the observed
$A^+$ value, and the resulting one to be entered in eq.(2) of \citet{ferraroetal2018}.
Because  $\Delta A^+$ $\approx$  0.0 dex at the Sun's galactocentric distance,
this procedure allows to compare dynamical ages of star clusters 
spread throughout the Milky Way as they were located at the position of the
Sun.

I also performed a linear least square fit between 
$A^+$,  $N_{relax}$ and $a$, as follows:

\begin{equation}
log(N_{relax}) = C_0 + C_1 \times A^+  + C_2 \times log(a /kpc),
\end{equation}

\noindent and obtained  $C_0$ = 1.52$\pm$0.18, $C_1$ = 4.57$\pm$0.41, and
$C_2$ =  -0.77$\pm$0.15, rms= 0.39, $\chi^2$= 0.16, and a
coefficient of determination R$^2$= 0.81. 
I obtained for eq.(2) a Spearman rank correlation coefficient 
of 0.87  and a Pearson correlation coefficient of 0.90, which represent an improvements over 
the values  of 0.82 and 0.85 obtained by \citet{ferraroetal2018}, respectively, 
for their eq. (2). 
For the sake fo the reader Figure~\ref{fig1} (right panel)
depicts the difference between $N_{relax}$ values calculated from \citet{ferraroetal2018}'s
relation and from eq.(2) as a function of $A^+$.



\end{document}